\documentclass{article}

\usepackage{microtype}
\usepackage{graphicx}
\usepackage{booktabs} 

\usepackage{hyperref}

\usepackage[preprint]{icml2024}

\usepackage{amsmath}
\usepackage{amssymb}
\usepackage{mathtools}
\usepackage{amsthm}

\usepackage[capitalize,noabbrev]{cleveref}

\theoremstyle{plain}

\theoremstyle{definition}

\theoremstyle{remark}

\usepackage[textsize=tiny]{todonotes}

\usepackage{multirow}
\usepackage{subcaption}
\usepackage{mdframed}

\icmltitlerunning{Why Solving Multi-agent Path Finding with Large Language Models has not Succeeded Yet}

\begin{document}

\twocolumn[
\icmltitle{Why Solving Multi-agent Path Finding \\with Large Language Models has not Succeeded Yet}



\icmlsetsymbol{equal}{*}

\begin{icmlauthorlist}
\icmlauthor{Weizhe Chen}{1}
\icmlauthor{Sven Koenig}{1}
\icmlauthor{Bistra Dilkina}{1}
\end{icmlauthorlist}

\icmlaffiliation{1}{University of Southern California}

\icmlcorrespondingauthor{Weizhe Chen}{weizhech@usc.edu}
\icmlcorrespondingauthor{Sven Koenig}{skoenig@usc.edu}
\icmlcorrespondingauthor{Bistra Dilkina}{dilkina@usc.edu}

\icmlkeywords{Machine Learning, ICML}

\vskip 0.3in
]



\printAffiliationsAndNotice{}  

\begin{abstract}
With the explosive influence caused by the success of large language models (LLM) like ChatGPT and GPT-4, there has been an extensive amount of recent work showing that foundation models can be used to solve a large variety of tasks. However, there is very limited work that shares insights on multi-agent planning. Multi-agent planning is different from other domains by combining the difficulty of multi-agent coordination and planning, and making it hard to leverage external tools to facilitate the reasoning needed. In this paper, we focus on the problem of multi-agent path finding (MAPF), which is also known as multi-robot route planning, and study the performance of solving MAPF with LLMs. We first show the motivating success on an empty room map without obstacles, then the failure to plan on the harder room map and maze map of the standard MAPF benchmark. We present our position on why directly solving MAPF with LLMs has not been successful yet, and we use various experiments to support our hypothesis. Based on our results, we discussed how researchers with different backgrounds could help with this problem from different perspectives.
\end{abstract}

\section{Introduction}

In the past year since ChatGPT came out, large language models (LLMs) have been shown to go beyond strictly language-related tasks like translation, and to be a powerful tool in all kinds of domains. Training with very rich and diverse datasets, LLMs incorporate a large variety of knowledge and do not require fine-tuning before generating good solutions in many real-world applications. Over time, recent studies have shown that combining LLMs with different ways of prompting can help solve problems that have some aspects of reasoning, including examples from logical problems \cite{yao2022react, liu2023dynamic} to controlling a robot dog without finetuning \cite{wang2023prompt}. 

Motivated by the success of LLMs, some recent works studied the performance of LLMs on multi-agent problems and showed LLM can also help multi-agent coordination \cite{chen2023scalable, agashe2023evaluating}. However, they barely cover multi-agent planning. In this paper, we consider the problem of multi-agent path finding (MAPF), also known as multi-agent route planning. MAPF is the problem of moving a group of agents from their respective start locations to their goal locations without collisions. MAPF can be used directly to formulate real-world applications like warehouse management \cite{DBLP:journals/ai/SharonSFS15, DBLP:journals/ral/HanY20}, swarm control \cite{DBLP:conf/socs/000100FKK20}, among others. In a typical warehouse scenario, around a thousand warehouse robots (agents) could be running simultaneously in one warehouse room (scenario), and each agent needs to plan its path, which could be as long as a hundred timesteps. Previous methods for MAPF can be classified into 1) classic methods like heuristic search and SAT \cite{DBLP:journals/ai/SharonSFS15, DBLP:journals/ral/HanY20}, and 2) learning-based approaches that mostly use reinforcement learning \cite{DBLP:journals/ral/SartorettiKSWKK19}. MAPF is a unique problem in multi-agent coordination in that coordination is required within the planning, so it is highly challenging to facilitate the reasoning of coordination by using LLMs to generate useful high-level guidance combined with a low-level path planner to ensure the solution is valid, an approach that will be parallel to this taken in the setting of overcooked \cite{agashe2023evaluating}. On the other hand, path planning is one of the easiest parts of planning, and constraints only concern obstacles and moving continuity. Because of these unique challenges in the MAPF problem, it is unclear how good LLMs will be at solving MAPF.

\begin{figure*}[t]
    \centering
    \includegraphics[width=0.9\textwidth]{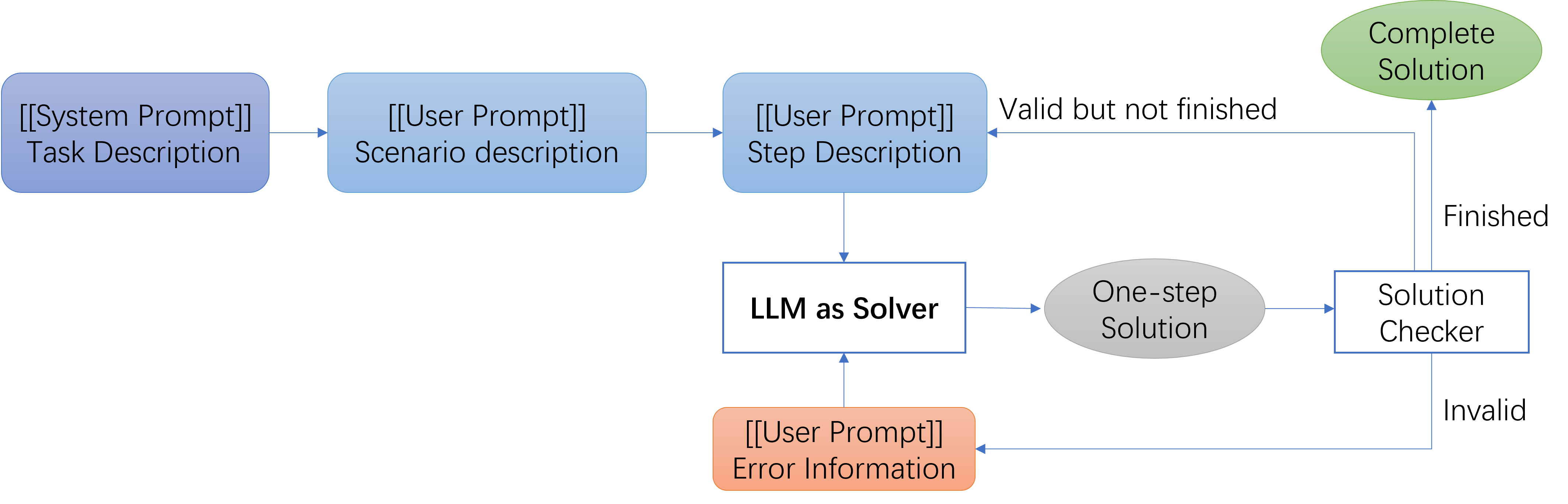}
    \caption{An illustration of our workflow.}
    \label{fig:workflow}
\end{figure*}

In this paper, we propose to use an LLM directly to solve MAPF without any additional heuristic guidance from solvers or any additional training. We give the scenario information to the LLM, and let the LLM recommend the actions for each agent every single timestep. We use a high-level checker that checks collisions between agents and obstacles to ensure the solution is valid in each step and provide error message to the LLM when violations occur, and iterate with the LLM until a valid solution is generated. We evaluate the performance of LLMs on the empty map, the room map, and the maze map, which are classic maps from the standard MAPF benchmark \cite{DBLP:conf/socs/SternSFK0WLA0KB19}. We show that LLMs can solve MAPF problems when the scenario is relatively easy and LLMs fail to generate any valid solution, no matter what the solution quality is when the scenario becomes harder. While LLMs are evolving extremely fast, and current failures may not apply to the next generation of LLMs, we elaborate on the current failing scenarios and give our position on three aspects as a breakdown of what kind of capability is currently missing in the LLM workflow, namely 1) the capability to understand the scenario, 2) the context length limit, and 3) the reasoning capability. We incorporate a list of experiments featuring various prominent prompt design alternatives, such as image-based and text-only inputs, and with or without single-step observation information, among others. Finally, we provide a discussion on the challenges of using LLMs for MAPF in the real world, and discuss how different researchers could contribute to the problem from different directions. We hope our work can serve as a building block for future research in foundation models for MAPF.

\section{LLMs for Multi-agent Path Finding}

\subsection{Preliminary}
The multi-agent path finding (MAPF) problem is the problem of finding a set of conflict-free paths for a set of agents in a known environment while minimizing their travel times. Specifically, in this paper, we consider the same problem as \cite{DBLP:conf/socs/SternSFK0WLA0KB19,DBLP:conf/aaai/0001CHSK22}, which is a four-connected grid map, where each agent is given a start cell and a goal cell. A scenario is defined as the combination of the description of the map, which is the size of the map, indication of which cells have obstacles, and the start cells and goal cells of each agent. At each timestep, an agent can move to an adjacent cell or stay in its current cell. A conflict happens if two agents end up in the same cell at the same timestep. Each agent remains at its goal cell after it arrives until all agents arrive at their goals. The objective is to minimize the makespan of the solution, i.e., minimize the time when the last agent arrives at its goal location. 

Large language models (LLMs) like ChatGPT, Claude, and Gemini have at least tens of billions of parameters and are trained with extremely large amounts of data collected on the internet. With the number of parameters exponentially larger than the degree of freedom in small problems, LLMs have the potential to solve some easy problems in reasoning with certain prompts \cite{DBLP:conf/nips/Wei0SBIXCLZ22}, and break down harder problems into smaller ones to get suboptimal solutions \cite{yang2023large}. When applying to MAPF, we hope LLMs can be an alternative model to the current MAPF reinforcement learning-based models without the need for any training. Unlike previous work with LLMs that solves the path-planning problem at the low level with existing heuristic solvers \cite{agashe2023evaluating}, MAPF is unique because it is extremely hard to decouple the high-level coordination from the path-planning process. The coordination needed is the deconflict that happens in each step and each step only, which forbids the use of a single-agent planner that cannot take coordination into account. And without the decoupling, we cannot leverage a path planner like previous work has done to reduce the difficulty in the reasoning.

\subsection{Methods}

\begin{figure}[t]
   \begin{mdframed}[linecolor=black, linewidth=1pt]
   \small
\textcolor{blue}{Agent 1 is currently in (0,2), and wants to go to (3,1).}

\textcolor{blue}{Agent 2 is currently in (1,3), and wants to go to (2,0).}

The map is as follows, where '@' denotes a cell with an obstacle that an agent cannot pass, and '.' denotes an empty cell that an agent can pass.

\textcolor{orange}{The bottom-left cell is (0,0) and the bottom-right cell is (31,0):}

\textcolor{orange}{....}

\textcolor{orange}{...@}

\textcolor{orange}{....}

\textcolor{orange}{.@..}

\textcolor{red}{In the next step:}

\textcolor{purple}{Agent 1 can move ['stay at (0, 2)', 'right to (1, 2)', 'up to (0, 3)', 'down to (0, 1)'].}

\textcolor{purple}{Agent 2 can move ['stay at (1, 3)', 'left to (0, 3)', 'right to (2, 3)', 'down to (1, 2)'].}

    \end{mdframed}
    \caption{An example of the user prompt for describing the scenario. Text in blue is a scenario-specific prompt, while text in orange is a map-specific prompt. In the experiments on the empty map, only the first blue paragraph will be provided, and all text starting from the black paragraph is removed because there are no obstacles. The text in purple is the single-step observation (SSO) information.}
    \label{fig:user_prompt_init}
\end{figure}

\begin{figure}[t]
   \begin{mdframed}[linecolor=black, linewidth=1pt]
   \small
[[Success]]

\textcolor{blue}{Good job. Keep moving. In the next step:}

\textcolor{blue}{Agent 1 can move ['stay at (0, 2)', 'right to (1, 2)', 'up to (0, 3)', 'down to (0, 1)'].}

\textcolor{blue}{Agent 2 can move ['stay at (1, 3)', 'left to (0, 3)', 'right to (2, 3)', 'down to (1, 2)'].}

[[Failure]]

\textcolor{red}{You are wrong. Agent (1,2), (4,5) are colliding with each other. Please correct the current step.}

\textcolor{orange}{You are wrong. Agent 2,4 is colliding with obstacles. Please correct the current step.}

    \end{mdframed}
    \caption{An example of the user prompt starting from the second step. While here we demonstrate a few options, only one of them, i.e., text in one color, will be provided to the LLM in one iteration.}
    \label{fig:user_prompt}
\end{figure}

In this paper, we focus on using LLMs to solve the MAPF problem directly.
However, it is quite obvious that current ML models cannot be perfect solvers in their first trial, and we introduce a high-level conflict checker to ensure the plan generated by the LLM is valid. We inform the LLM about the mistake in the solution, if any. Unlike some previous works \cite{yao2022react, yang2023large}, our checker is not another LLM both because it is extremely easy and efficient to detect collisions by a rule-based detector in linear time, and also because the LLM fails to always correctly identify the collisions. In this checker, we not only check for agent-to-agent collisions but also correct the LLM for any collision with the fixed obstacles. We do not provide any additional guidance on how the LLM should deconflict the solution but just describe the conflicts.

Following the common practice of LLMs, we build our workflow shown in Fig.~\ref{fig:workflow}. As existing learning-based approaches \cite{DBLP:journals/ral/DamaniLWS21}, we give LLMs stepwise local information and let the LLM choose the actions of agents step-by-step. This step-by-step (SBS) generation is different from the popular chain-of-thoughts idea \cite{DBLP:conf/nips/Wei0SBIXCLZ22} used in LLMs by not introducing more intermediate reasoning processes in the generation. Instead, it breaks down the whole planning task into smaller single-step tasks, so the LLM does not need to be fully correct before we can use some results from it. We start by giving the LLM the system prompt of becoming a solver for the MAPF problem. This part is fixed within each map, and different in the map description part in different maps. Then, we start the user prompt by providing the scenario information, which includes where the obstacles are, as well as the start location and the goal locations of each agent. We found that providing LLMs with specific local information about viable actions for the immediate next step significantly helps them avoid collisions with static obstacles, so we provide a single-step observation (SSO) in the prompt by default. See Figure \ref{fig:user_prompt_init} for an example of the scenario description prompt. Then we start to read the output of the LLM, use the checker to determine whether the output of the LLM is valid, and either output the errors to let the LLM correct on the current step, or tell the LLM to move on to the next step with the information that is specifically for the next step. See Figure \ref{fig:user_prompt} for an example of the iteration-wise user prompts. To address the total token limit, we leverage the fact that MAPF is a Markov decision process where each state is independent of its previous states, and we restart the prompt from scratch, i.e., treat the current agent locations as their starting point, whenever we got a rate limit error.

\subsection{Experiment Results on MAPF Benchmark}

\begin{figure}[t]
    \centering
    \begin{subfigure}[b]{0.3\linewidth}
    \centering
    \includegraphics[width=0.75\columnwidth]{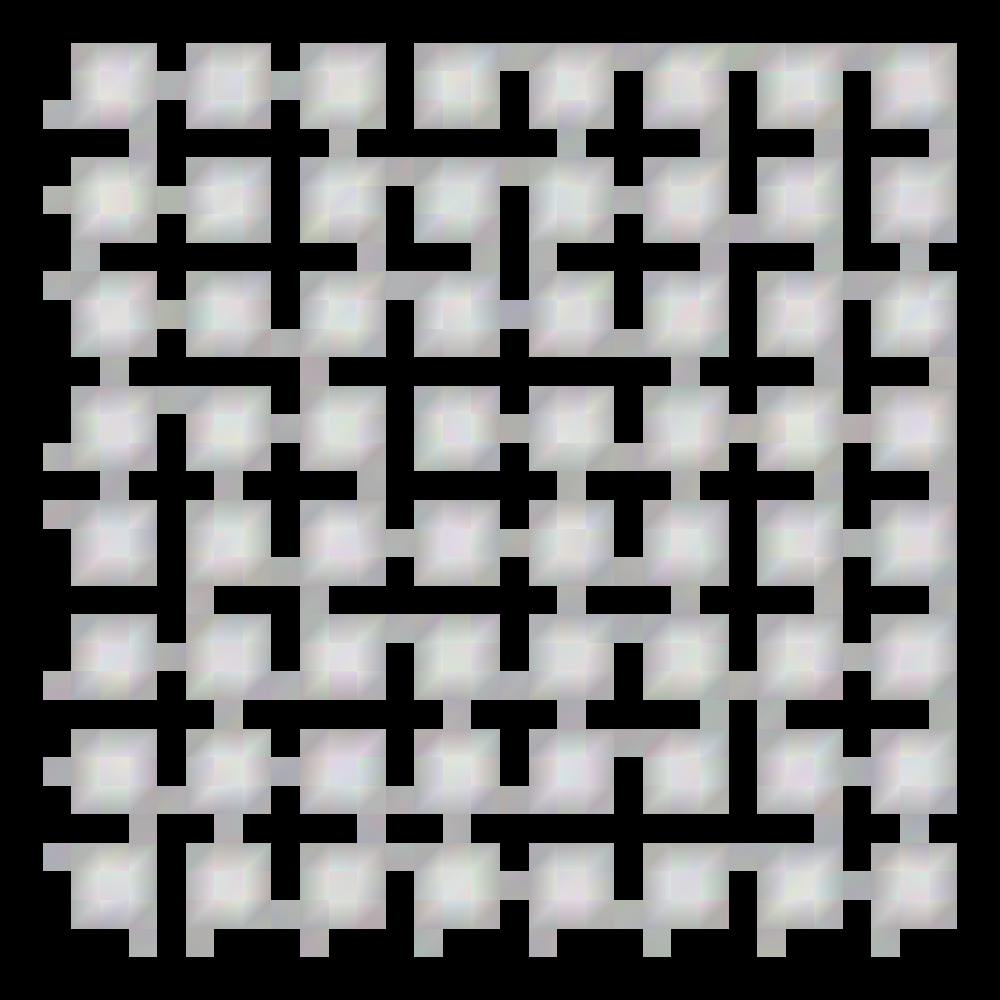}
    \end{subfigure}
    \centering
    \begin{subfigure}[b]{0.3\linewidth}
    \centering
    \includegraphics[width=0.75\columnwidth]{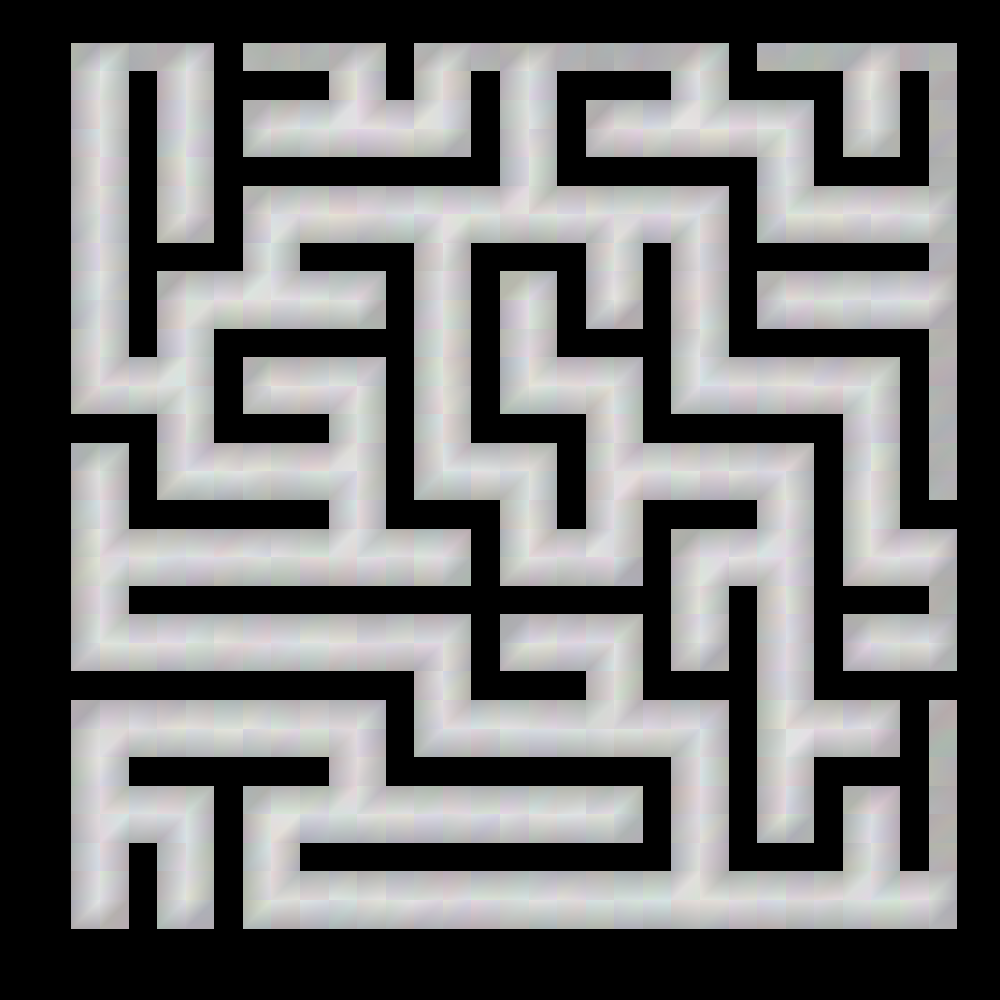}
    \end{subfigure}
    \caption{The \textsc{room}-32-32-4 map (left) and the \textsc{maze}-32-32-2 map (right). The picture is vertically flipped to match the common knowledge that higher positions on the vertical axis indicate greater values.}
    \label{fig:room_map}
\end{figure}

\subsubsection{General Experiment Settings}
\label{sec:failing_criteria}
While we have introduced the high-level conflict detector, we define a solution generation to be successful if it does not fail because of any of the following:
\begin{enumerate}
    \item Fails to generate a plan whose number of steps is at most 3 times longer than the optimal plan.
    \item Fails 5 consecutive times in a single step after we provided the first round of feedback, specifying whether the current solution is correct or not.
\end{enumerate}

Note that we do not consider any token limit error because allowing no more than 5 times in a single step guarantees that the message length in one step is within the token limit, and we already consider restarting at each step if the token limit is too long in our method.

By default, we are using the GPT-4-1106-preview model, also known as the GPT-4-turbo model, which is the state-of-the-art LLM that is publicly available. We set the temperature as $1.0$ and the seed as 42 \footnote{Note that we are testing using the API, where the behavior could be slightly different from the web version of GPT-4.}. We show an example of our user prompt in Fig.~\ref{fig:user_prompt_init} and Fig.~\ref{fig:user_prompt}, and an example of the complete system prompt in the appendix. We will describe how each variant will change from this default prompt. 

Each setting in the experiment is tested on 5 different scenarios in the standard MAPF benchmark, each different in terms of the start and goal location combinations of the agents on a given map.

\subsubsection{Results}
While it is unclear how well LLMs can solve MAPF problems, we start by showing that LLMs can be a promising solver in easy domains. We test the performance of LLMs on the \textsc{empty} map from the MAPF benchmark \cite{DBLP:conf/socs/SternSFK0WLA0KB19}, which has no obstacles at all and only requires the agents to not collide with each other. The results are shown in Table.~\ref{table:benchmark}. We observe that for smaller-scale problems that has fewer agents, GPT-4 can successfully generate valid solutions. However, GPT-4 fails to generate valid solutions when there are more agents on the same map. 

It has been shown that a step-by-step generation is better in many reasoning-related tasks, since it can guide the generated plan to create the chain of thoughts as a human rather than guessing a solution and then checking whether they are correct \cite{DBLP:conf/nips/Wei0SBIXCLZ22}. However, it is not well-studied whether a step-by-step generation is also helpful in the planning since a complete plan will also require to be given in a step-by-step (SBS) way, and the LLM models need to decide every single step anyway. Therefore, we study a variant that generates the whole plan in one shot (OS) with the help of chain-of-thought prompts and show the comparison in Table.~\ref{table:benchmark}. We observe that one-shot (OS) generation fails to generate solutions even in scenarios with 4 agents. This is because the whole plan is quite long, and the total number of possible failures is the product of possible failure at each step, making it inefficient to find a valid solution.

Motivated by the success on the \textsc{empty} map, we tested the same workflow on the \textsc{room}-32-32-1 map and the \textsc{maze}-32-32-2 map, which is shown in Fig.~\ref{fig:room_map}. The results are shown in Table.~\ref{table:benchmark}. In this case, we found that LLMs fail to generate solutions when the number of agents is already quite small, and that LLMs completely fail on the \textsc{maze}-32-32- map even with only 2 agents. 

\begin{table}[t]
\small
\centering
\begin{tabular}{llll}
\hline
\multirow{2}{*}{Map} & \multirow{2}{*}{n} & \multicolumn{2}{l}{Success Rate (\%)} \\ \cline{3-4} 
                     &                    & OS             & SBS             \\ \hline
                     &            2        &       10         &      100           \\
Empty                     &        4            &       0         &     100            \\
                     &       8             &               0 &       100          \\
                     &       16             &          0      &       60          \\\hline
                     &            2        &       10         &      100           \\
Room                     &        4            &       0         &     80            \\
                     &       8             &      0            &      20           \\\hline
Maze                &   2 & 0 & 0 \\\hline
\end{tabular}
\caption{The success rate of solving MAPF with LLM on \textsc{empty}-8-8, \textsc{room}-32-32-4 and \textsc{maze} maps with varying numbers of agents (n). }
\label{table:benchmark}
\end{table}

\section{Cause of Failures}
Because LLMs are evolving rapidly, it is unclear how long LLMs will still fail. Although our methods do not combine any external tools like low-level planning or external memory \cite{agashe2023evaluating}, which could make LLMs much stronger, it is important for us to break down the failures and identify the cause of the failures so we can strategically choose the correct tools to overcome them in the future. We found that the cause of failing in MAPF is different from the cause of failing in general planning tasks where the generated plan is invalid, usually by breaking requisite constraints \cite{valmeekam2023planning}. In path planning, the only constraint is that one agent needs to move to an adjacent cell and cannot jump across the map. This is a very easy task for LLM to follow and thus, in our experiments, we never see the LLM failing due to that constraint, even if the only instruction about this constraint is "in a 4-connected grid". We observe that LLMs fail only once with 16 agents in the empty map because of consecutive failures in a single step, and all other times they fail because they use many more steps than they should. More specifically, we discovered that 77\% of the failures occurred because the LLM agents began to oscillate in a specific area of the map, while the remaining failures were due to excessively long detours.

Based on our observations, we break down the reasons into three deeper parts from the perspective of natural language models: limitation from the model, understanding, and reasoning. For simplicity, we first discuss the lack of reasoning capability, which is common in many tasks.

\subsection{Reasoning Capability}

\begin{figure}
\centering
    \begin{subfigure}[b]{0.35\linewidth}
    \centering
    \includegraphics[width=0.85\columnwidth]{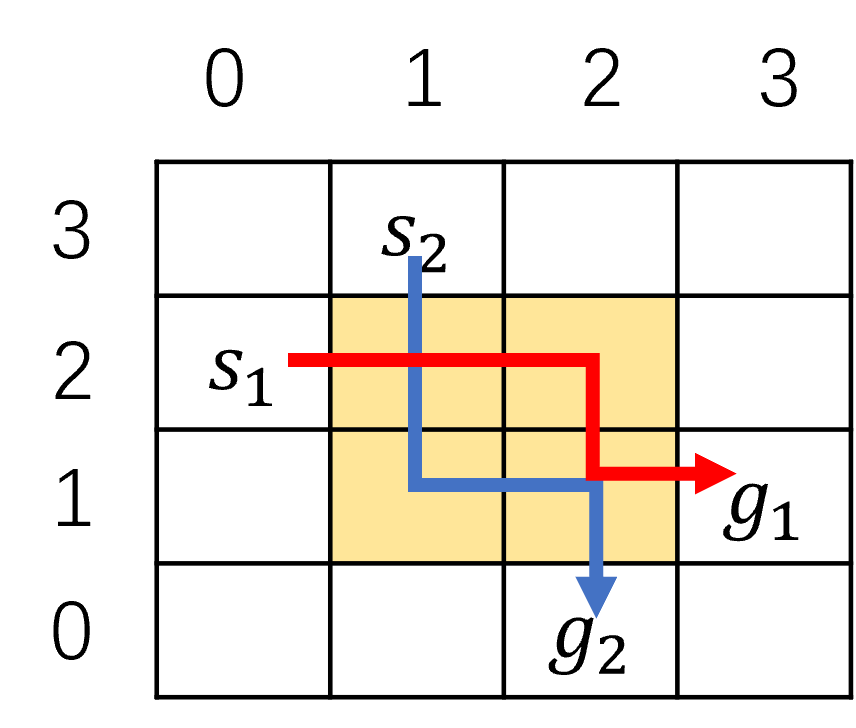}
    \caption{}
    \label{fig:sym_breaking_1}
    \end{subfigure}
    \centering
    \begin{subfigure}[b]{0.35\linewidth}
        \centering
    \includegraphics[width=0.85\columnwidth]{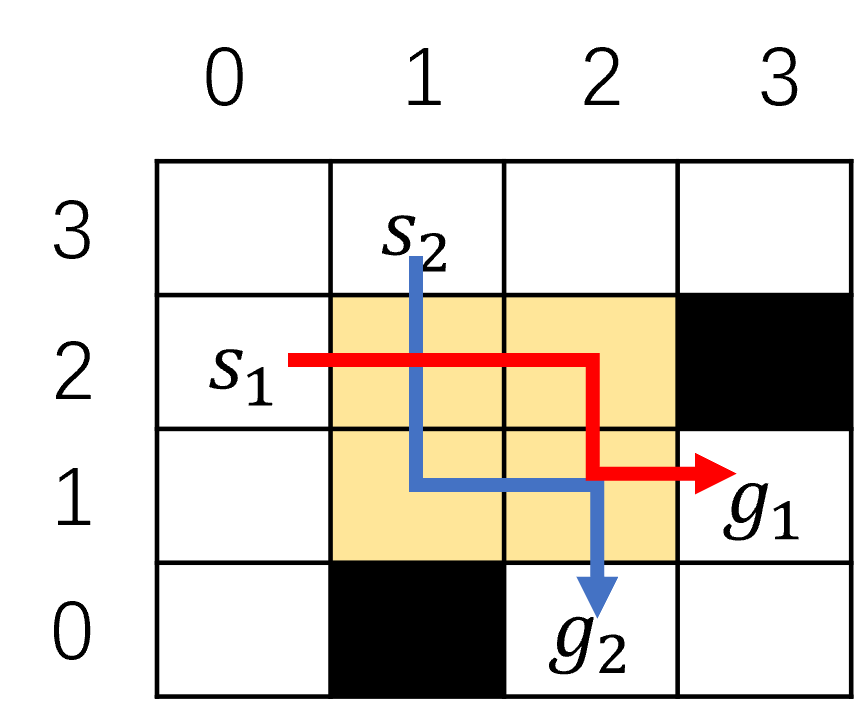}
    \caption{}
    \label{fig:sym_breaking_2}
    \end{subfigure}
    \caption{Two examples of symmetry breaking examples, originally from \citet{li2019symmetry}. In \ref{fig:sym_breaking_1}, every pair of shortest path will collide with each other. In \ref{fig:sym_breaking_2}, every pair of shortest path will collide with each other in a fixed cell (1,2).}
    \label{fig:sym_breaking_fig}
\end{figure}

\begin{figure}[!ht]
    \centering
    \begin{mdframed}[linecolor=black, linewidth=1pt]
    \small
   Step 1: \newline
Agent 1: \newline
Current: (0,2) \newline
Goal: (3,1) \newline
Move: Right to (1,2) \newline
\newline
Agent 2: \newline
Current: (1,3) \newline
Goal: (2,0) \newline
Move: Down to (1,2) \newline
\newline
After the move, the coordinates of the agents are: \newline
Agent 1: (1,2) \newline
Agent 2: (1,2) \newline
\newline
There is a collision between Agent 1 and Agent 2. I will regenerate another solution immediately. \newline
\newline
Step 1 (Regenerated): \newline
Agent 1: \newline
Current: (0,2) \newline
Goal: (3,1) \newline
Move: Down to (0,1) \newline
\newline
Agent 2: \newline
Current: (1,3) \newline
Goal: (2,0) \newline
Move: Left to (0,3) \newline
\newline
After the move, the coordinates of the agents are: \newline
Agent 1: (0,1) \newline
Agent 2: (0,3) \newline
\newline
There is no collision. Here is the validated solution: \newline
\newline
Agent 1: (0,1) \newline
Agent 2: (0,3)
    \end{mdframed}
    \caption{The output from GPT-4-turbo for the symmetry breaking example in Fig.~\ref{fig:sym_breaking_2}.}
    \label{fig:output_sym}
\end{figure}

\begin{table}[t]
\small
\centering
\begin{tabular}{@{}llllll@{}}
\toprule
\multirow{2}{*}{Model} & \multirow{2}{*}{n} & \multicolumn{2}{l}{Success Rate} & \multicolumn{2}{l}{Avg. Iterations} \\ \cmidrule(l){3-6} 
                       &                    & GO            & GO+SSO           & GO              & GO+SSO             \\ \midrule
                       & 2                  & 80            & 100              & 2.7               &      1.6             \\
GPT-4-8K                & 4                  & 20            & 60               &       3.0          &     2.3               \\
                       & 8                  & 0             & 0                & N/A             & N/A                \\ \midrule
                       & 2                  & 100           & 100              & 2.1               & 1.2                  \\
GPT-4-128K       & 4                  & 60            & 80               &          2.7       &         1.4           \\
                       & 8                  & 0             & 20               & N/A             & 2.4                \\ \bottomrule
\end{tabular}
\caption{The success rates and average iterations per step used until proceed to next step in success scenarios for GPT-4 and GPT-4 Turbo, whose token limits are 8K and 128K respectively, on the room-32-32-4 map with different number of agents(n). }
\label{table:context_length}
\end{table}

In this paper, we focus on optimizing the makespan of the MAPF solution, and making the agents prefer to wait rather than detour in small scenarios that do not lead to endless waiting. We calculate the average ratio of makespan to the maximum length among the single agent shortest paths (a standard normalization practice in MAPF as this provides a valid lower bound on the optimal MAPF solution), and in scenarios that are successfully solved, this average is 1.5. This means that LLMs can mostly succeed in scenarios that do not need a lot of waiting, and in most steps, they only need to go in the two directions aligned with the goal's direction. LLM fails in other scenarios because they need path finding in complex environments. A simple example of the failure of path finding is shown in Fig.~\ref{fig:sym_breaking_fig}, where we have the classic symmetry-breaking example. We provide the input of the scenario information and the output of the first step in Fig.~\ref{fig:user_prompt} and Fig.~\ref{fig:output_sym}. We observe that while waiting for one step can clearly lead to the optimal solution that has a makespan of 5, the LLM chose to move agent 2 to the left and resulting in a total makespan of 6. This example shows that LLM does not understand what makespan is and how a solution is optimized for makespan. 

On the other hand, when we look at the average number of iterations for prompts with both global obstacle observation in the first user prompt and single-step obstacle observations(GO+SSO) in Table.~\ref{table:context_length}, it is still not very close to $1$ which means no iterations at all. This is because even if we have explicitly let the LLM deconflict the agents by listing all the coordinates of agents, it does not have the capability to check the answers itself without the use of external tools. More specifically, LLM cannot tell whether a list of tuples of numbers has any duplicates. This is similar to the fact that LLM cannot sort a list of numbers without step-by-step instructions, which we do not allow the LLMs to do because we still do not want the LLM to use too many tokens in a single step. We also observe that the GPT-4-8k model takes more iterations than the latest GPT-4-turbo model. This could either contribute to forgetting earlier information or to the improved capacity of the new model.

This lack of reasoning capabilities is usually solved with tool use in other domains, but MAPF itself is hard because MAPF requires the capability of path finding in a complex map and avoiding collisions between all pairs of agents strategically and efficiently. The coordination between agents is required in each step, and each step only. If we let the tool, in this case, a single-agent planner like A*, include the de-conflicting, the problem is already solved, while LLM did no help. On the other hand, the tools must know what kind of deconflict they need to do to add those constraints into the heuristic search algorithm. This paradox is making LLM with tool use in MAPF very hard.

To potentially increase the reasoning capability, we have also tried a list of tricks. These kinds of tricks include breaking the big paragraph into bullet-point style instruction, removing some 'useless' instruction, adding a whole example for in-context learning, and using the theory of mind to guess other agent's intuitions \cite{DBLP:conf/emnlp/LiCSCHLS23}. These tricks did not show any help in any scenarios, and even greatly harmed the performance of the GPT-4-8k model.

\subsection{Context Length Limit}

Because the underlying architecture of popular LLMs is transformers, which further rely on self-attention, a longer context in the input will significantly increase the computing complexity in the process. Therefore, in their training process, they set a maximum limit of tokens on the input and trained specifically on them. Current large language models released will also provide a context length limit, which is as long as 200K tokens, and any request with a longer context length will be rejected. This is a long enough length for many tasks and even able to read a textbook, and many users are satisfied with the length. However, recent studies have shown that long models like GPT4-turbo-128K are not a model whose capacity in 8K length also works when given a 128K-tokens input \footnote{https://github.com/gkamradt/LLMTest\_NeedleInAHaystack}. What is even worse is that in the real-world applications of MAPF, the total number of agents running in the environment could be up to a few thousand. Even in the ideal case that our prompt will grow linearly to the number of agents, an environment with a thousand agents will require a total context length of $250 (tokens /agent\ step) * 50 (steps) * 1000 (agents)=5M$ tokens, which is much bigger than the current limit. Because the token limit is strict, we have introduced the restart mechanism in our methods, and in our current experiment scale, we have to restart the LLM around every 7-10 steps in large or complex scenarios. We also observed that the number of tokens used grows non-linearly with regard to the number of agents due to the increasing number of iterations, and we have put the numbers in the appendix. 

In table.~\ref{table:context_length}, we show our results on the \textsc{room} map, and test with different GPT-4 models with different context lengths. We found that when the context length limit grows, the success rate also increases. This is especially helpful when the single-step observation information that tells what valid actions each agent has is not provided. The failures here are not direct failures by exceeding the context length limit, but from the forgetting in restart caused by reaching the context length limits. Because of these restarts, our models often completely forget that they have been to certain locations, and then go back and forth in certain areas with dead ends like the center room in the \textsc{room} map.

\subsection{Understanding Obstacle Locations}

\begin{table}[t]
\small
\centering
\begin{tabular}{lllll}
\hline
n\textbackslash{}Success Rate & MM  & TOO & TOM \\ \midrule
2                             & 100 & 100 & 100 \\
4                             & 20  & 60  & 80  \\
8                             & 0   & 0   & 20  \\ \bottomrule
\end{tabular}
\caption{The success rate of different ways of inputting the map information to LLM on \textsc{Room}-32-32-4 map with different number of agents (n). }
\label{table:abla_mm}
\end{table}

\begin{figure}
    \centering
    \includegraphics[width=0.85\columnwidth]{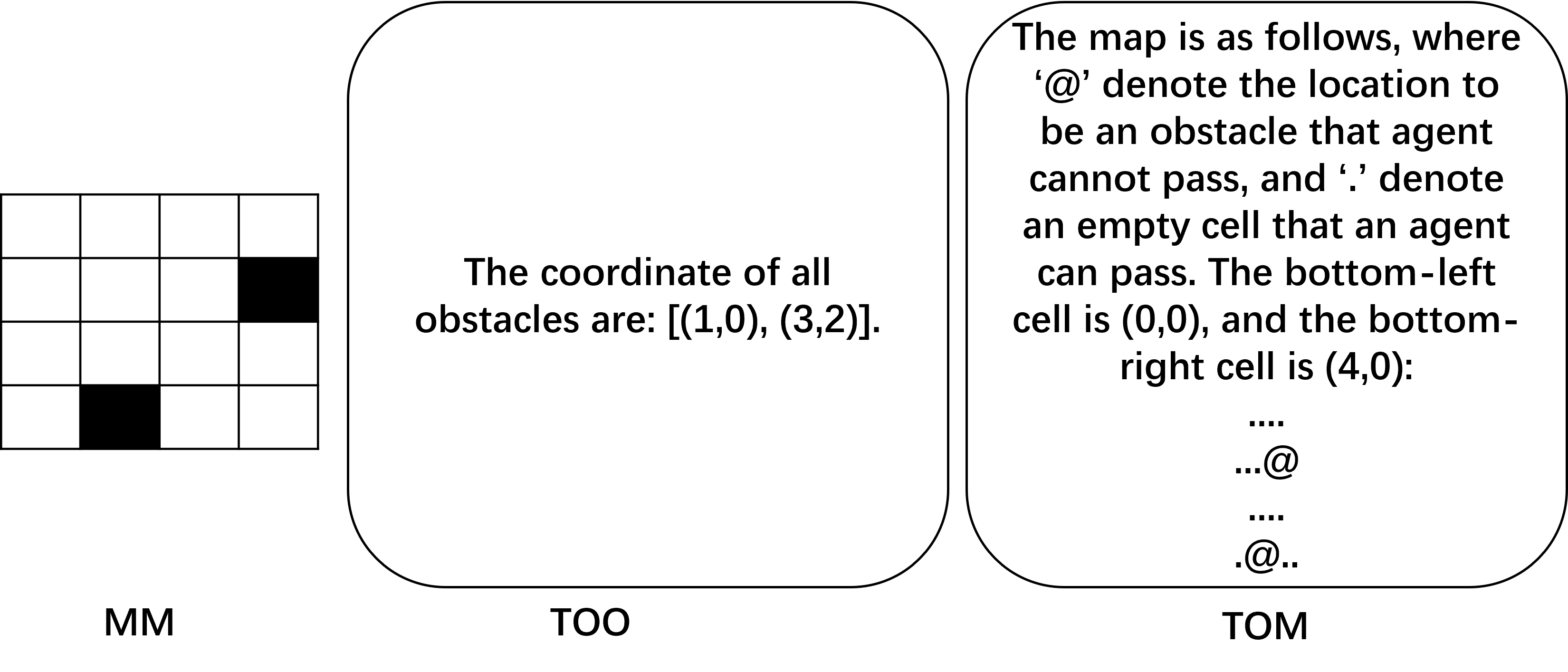}
    \caption{An illustration of the difference between multimodal input (MM), text that describes the whole map (TOM), and text that describes the obstacles (TOO).}
    \label{fig:abla_mm}
\end{figure}

MAPF scenarios can be broken down into a pair of start location and goal locations, together with the map information. While understanding the coordinate version of start goal location pairs is relatively easy, understanding the map information is hard. Trained with publicly available text, LLM learns what a specific map means by finding similar contexts online, which mostly comes from other related fields in planning, like solving a maze. However, one problem is that people barely provide any such information online since people have the common knowledge of what to do with a map with code and preprocess the map information in the code rather than explicitly provide the set of where the obstacles are as the original input. Therefore, this leads to a lack of training data with related context about the information in the pictures or in a symbol-based input.

Because of this lack of information, our main prompt included guidance on what action could be taken in the current step to increase the success rate a little bit. In Table.~\ref{table:context_length}, we provided the results of global observation (GO) only and the version with single-step observation (SSO) together. We discover that while global observation can also provide a good win rate with the help of our high-level checkers, it increases the average iterations in each step, and also fails in cases that could be successful with the help of SSO, which is killed by using much more total number of steps than it should. In general, LLMs fail in medium complexity maps like \textsc{room}, and completely fail in much harder maps like \textsc{maze}, even with the help of SSO. 

In all the experiments above, we use pure-text-based input because of the popularity and simplicity of pure-text input. However, this was not the case before: In earlier work of reinforcement learning for MAPF, people transformed the input into image-based input within the field of view and used neural networks like convolutional neural network (CNN) to read the information.

Therefore, as shown in Fig.~\ref{fig:abla_mm}, we propose three different methods, which all share the same prompt of describing the problem definition and use different versions of the prompt describing the scenario: 
\begin{enumerate}
    \item A multimodal-based method MM where we use an image to give the picture of the map, and text to describe the problem and the scenario. 
    \item A text-only method that describes the map by listing all the coordinates of the obstacles, named TOO.
    \item A text-only method that describes the map by feeding the text version of the map with '@' denoting the obstacles and '.' denoting an empty space, named TOM.
\end{enumerate}

We test the performance of these three variants, and show the results in Table.~\ref{table:abla_mm}. We found that with an image-based input, the results are even worse. For example, large models can only have a high-level idea that there is a wall in the middle area, rather than knowing there is a wall exactly from $(3,10)$ to $(20, 10)$. The failure of TOO further indicates that LLMs do not know how to combine high-level intuitions of where obstacles are with concrete reasoning during path planning. Overall, in complex environments such as mazes and real-world city maps, LLMs cannot offer any form of guidance, not necessarily the specific action recommendations tested in this paper, until they can accurately comprehend the relationship between obstacles and the paths of agents.

\section{Discussion}
As we have shown in the previous section, there are three aspects that mainly limit the current capability of solving MAPF with LLMs. While there have been many works in the NLP that could solve one of them, e.g., extending the context length limit to up to 4M tokens or using a decentralized control, those methods will harm the reasoning capability and, therefore, are not preferable in our case.  


We would like to note that the methods in this paper are not designed to either outperform the state-of-the-art classic solvers or surpass the performance of learning-based solvers. Instead, the goal is to demonstrate that LLMs can solve small problems simply through prompting and discussing what is stopping them from solving larger ones.

Moving on, we believe the three reasons for failures can each independently relate to different research directions. Improving capability in a long context is already a popular direction in NLP, and there are also many papers looking into improving the general reasoning capability. We believe that much ongoing research has the opportunity to improve the performance of LLMs on MAPF, and we are pleased to have more researchers test this unique problem. Also, the recent development of building foundation models that combine language and image generation could be potentially helpful for the reasoning in MAPF. Image generation models like stable diffusion \cite{rombach2022high} have shown to be a strong tool in single agent planning\cite{janner2022planning}. While these diffusion models are not designed for multi-agent planning and are even hard to apply to MAPF, strategically using them could partially help the lack of reasoning capability. Regarding the understanding of the obstacles, we believe more annotated data are needed to help foundation models connect the abstract description to concrete coordinate-level knowledge of where obstacles are. Future research could finetune using their data, or at least open their data to the general public, and let the next version of released models from industry companies include such data by automatically getting such data online. Hopefully, such data could help foundation models able to know position encodings are specifically more important in problems related to planning. However, a challenge for fine-tuning a customized model is that popular open-source models have smaller context length, like 4K in LLAMA-2, and 32K for Mixtral 8*7b, which will limit their performance on MAPF.

Besides, although the success rate is one of the most important factors in measuring the performance of a solver, there are also other obstacles that need to be solved before using LLMs in real-world scenarios. First of all, success does not indicate anything about the solution quality. In our experiment, we found that the success scenarios usually come from easier scenarios that agents mostly only need to follow their direct shortest path. And when the planning is getting harder, the success rate is much worse. To improve the reasoning capability, future researchers and users could choose to make a dataset of scenarios and their good solutions publicly available on the internet, so when training with newer data, this specific problem could be directly included in the training set. Furthermore, the current workflow does not include any heuristics or tool use. While we are unclear on how such things should be included, this will be a very challenging but meaningful direction. For potential researchers going in this direction, we want to remind them to keep looking at the consistency of the deconflict of agents, which could be seen as moving obstacles, and the deconflict of the fixed obstacles on the map. A successful method should consider both difficulties in a similar manner.

On the other hand, latency is always a problem for LLMs when used in real-world production. If one wants to use their own models, which is currently ~500 tokens per second after using state-of-the-art system-wise optimization like vLLM \cite{kwon2023efficient} for LLAMA-2 \cite{DBLP:journals/corr/abs-2307-09288}. However, this speed is not enough as a typical round of output will have 1500 tokens, and it may take a few rounds before the LLM generates a valid solution. If one wants to use models provided through API like GPT-4 from OpenAI, the latency of the models will depend on a list of factors like network connection and server availability. In our paper, we typically need to wait around 15-30 seconds for one step to get completed, where less than 0.1 seconds are used for local processing.

\section{Related Works}

\subsection{LLM for Reasoning}
Since the release of GPT-3, researchers have studied the effect of how they are performing on diverse sets of problems. Starting with failure in general tasks\cite{DBLP:journals/corr/abs-2206-10498}, ideas like chain of thought \cite{DBLP:conf/nips/Wei0SBIXCLZ22}, in context learning \cite{ye2023context} have been shown to significantly improve the performance of LLMs in different tasks related to reasoning\cite{fu2022complexity, shum2023automatic}. A recent line of work has been introduced to use natural language as feedback in the process to introduce iterations and give LLMs more than one chance to generate correct solutions, which has been shown to significantly help in code generation \cite{shinn2023reflexion, chen2023teaching, olausson2023demystifying}, reasoning \cite{yao2022react, yang2023large, shinn2023reflexion}, and reducing harmful information in LLMs \cite{ganguli2023capacity}. There is also a very popular line of work augmenting LLMs with external memory, belief, and tools. For example, \citet{liu2023llm+} proposed to use LLMs to generate a PDDL description of the problem in classic planning benchmark domains like blocksworld and thermes. \citet{mao2023language} used external memory to match the previous experience and shows that the underlying knowledge in LLMs can make a novel tool that is more powerful than before in autonomous driving. In this paper, we use our high-level checker as the only tool in the process, and we hope the reasons for failure could later help researchers choose external tools.

\subsection{LLM for Multi-agent Systems}
While there is extensive research on using LLMs for many different problems, there is little work that addresses the problem of LLM for multi-agent systems in the beginning. This is typically due to the context length limit that generally applies when the total number of agents is not too small. As time proceeds, there has been a list of works that promote research in social behavior by creating multi-agent environment powered by a lot of LLM agents \cite{tan2023language, li2023camel}, create dialogue-based games with LLM \cite{schlangen2023dialogue}. Recent work has shown that dialogue between multiple agents can help accuracy and remove factual errors \cite{DBLP:journals/corr/abs-2305-14325, DBLP:journals/corr/abs-2307-05300}. When it comes to solving problems related to multi-agent system, there are a few works that are more in the multi-robot system. \citet{zhang2023building} and \citet{mandi2023roco} are two works that first extend LLMs to the domain, but they are limited to the scale of two to three agents. \citet{chen2023scalable} studied whether using a more decentralized controller can help to solve the context length limit problem when the number of agents is higher. \citet{agashe2023evaluating} evaluate the overall multi-agent coordination capability in three different domains and show that with proper tools, LLMs can even outperform MARL algorithms. While they succeed in their domains, in this paper, we specifically look at the problem of MAPF, and we are showing the cause of failure in the MAPF problem because of its unique challenges in MAPF. 

\subsection{Multi-agent Path Finding}
Multi-agent path finding (MAPF) is a problem that has won much attention in recent years because of its close relationship to real-world applications like autonomous warehouses. It has previously been solved with more classic methods like heuristic search algorithms \cite{DBLP:journals/ai/SharonSFS15}, rule-based algorithms \cite{DBLP:journals/ral/HanY20}, and reduction-based algorithms \cite{DBLP:conf/ecai/SurynekFSB16}. While learning-based approaches have not yet outperformed classic approaches, they have also begun to win a lot more focus for their fast inference time and generalizability. PRIMAL \cite{DBLP:journals/ral/SartorettiKSWKK19} proposed to learn a policy for MAPF using a combination of reinforcement learning and imitation learning. Following that, a group of works proposed a diverse set of methods from building curriculum \cite{DBLP:journals/ral/DamaniLWS21} to follow guidance from classic methods in each step \cite{skrynnik2023learn}. In this paper, we focus on using LLM, which is a neural-network-based approach but is not specifically fine-tuned for MAPF problems.  

\section{Conclusion}

In this paper, we investigate the use of LLMs for solving the MAPF problem. We first show that LLMs could work in easy scenarios like a small number of agents in the empty map, while being unable to stably generate valid solutions when the problem gets harder. To make our work applicable to future models, we elaborate on the failures and show that they can be broken down into the problem of the context length limit of the model, the capability to understand and remember where obstacles are, and the capability of planning. We used extensive experiments to support our breakdown, and provided discussions on how the difficulties could be addressed by people with different backgrounds. 

\section{Impact Statement}
This paper presents work whose goal is to express our position on how to advance the field of Machine Learning and Multi-agent Path Finding. There are many potential societal consequences of our work, none of which we feel must be specifically highlighted here.

\bibliography{main}
\bibliographystyle{icml2024}

\newpage
\appendix
\onecolumn
\section{Complete System Prompt}

\begin{figure*}[!ht]
   \begin{mdframed}[linecolor=black, linewidth=1pt]
   \small
You are a decision-maker in a warehouse system that is running as a 4-connected grid-based multi-agent path finding system of \textcolor{blue}{[[Map Description]]}. You will be provided with the current scenario in which you need to choose actions for each agent, i.e., the detailed information that provides where the obstacles are, together with each agent's current location and goal locations. The objective is to minimize the time of the agent who arrives at its goal location at the latest while keeping no agents from colliding with each other at any timestep. You only need to consider collision between two agents located in the same cell after each move. On each timestep, you could choose one action from moving up, moving down, moving left, moving right, and staying without moving. You should try to avoid collisions not only in the current timestep but also look ahead to a few timesteps if it is important to avoid future collisions. Remember that to avoid a collision, normally, only one agent needs to change their action from the original action. Nevertheless, your job is not to generate a whole path but to do it step-by-step, and the information on whether a specific action is valid or not is provided in each single step. Remember that the given action is based on the obstacles, so the available action will not change if the agent chooses to stay at the same location. You only need to give a single-step action for each agent to follow; do not give more steps until I tell you to. 
First, give your solution in the same format as follows, with some explanation if the agent is not moving or moving in a direction that is not towards its goal. (This is not the actual scenario, but an example of the desired output format. You will be provided with the actual scenario later):

Step 1:

Agent 1:

Current: (1,0)

Goal: (1,0)

Move: Stay, as it has already reached its goal.
\newline

Agent 2:

Current: (5,4)

Goal: (5,6)

Move: Up to (5,5)
\newline

Agent 3:

Current: (0,7)

Goal: (6,4)

Move: Right to (1,7)
\newline

Then, validate your solution by listing all the coordinates of the agents after the move and check if there is any pair of agents that have the same coordinates. If there is a collision by mistake, regenerate another solution immediately and validate again by listing the coordinates of all the agents until you get a collision-free solution. Finally, end your output with your validated solution in a new paragraph with the format of a sequence of : [[Agent\_id]]: [[Coordinate]].
    \end{mdframed}
    \caption{An example of the system prompt to specify the MAPF system and the objective. The text highlight in blue will be replaced by map description. For example, in room-32-32-4, the prompt will be 'room-like map with size 32*32'. }
    \label{fig:system_prompt}
\end{figure*}

In Fig.~\ref{fig:system_prompt}, we give our system prompt. Between different map, the only thing changed is the text highlighted in blue, and all others part are the same.

\section{Token Length Grows}

\begin{figure}
    \centering
    \includegraphics[width=0.6\columnwidth]{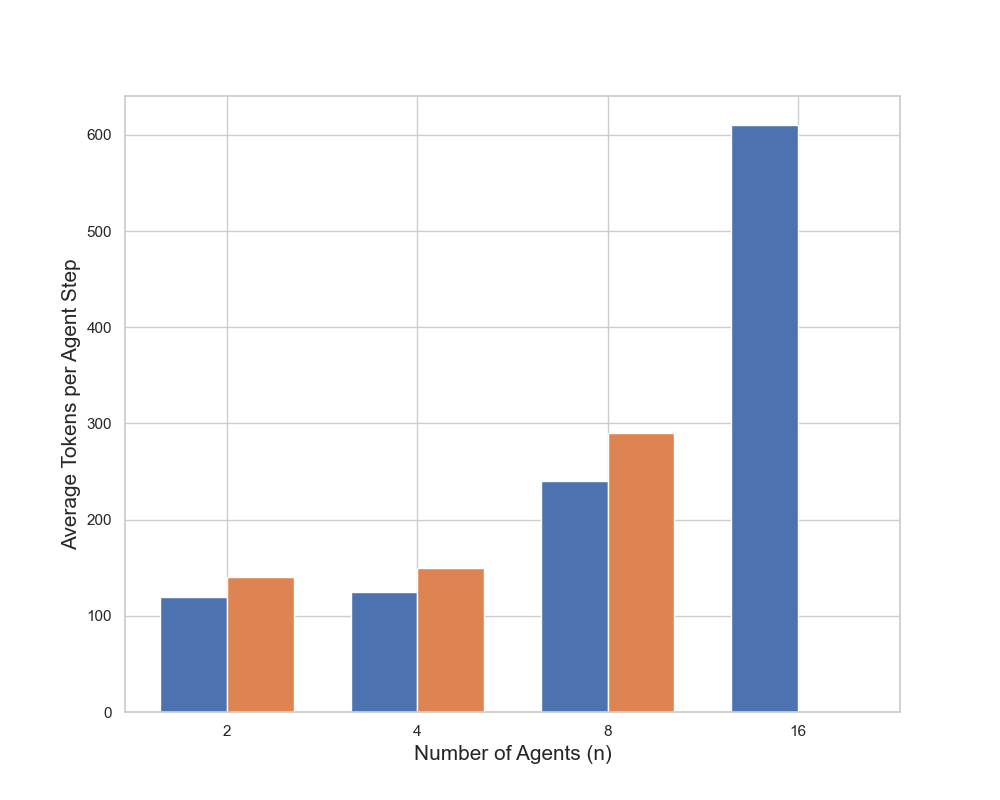}
    \caption{Average token length per agent step on success scenarios.}
    \label{fig:token_growth}
\end{figure}

In Fig.~\ref{fig:token_growth}, we show the average prompt length per agent per step, which includes both the input and the output. We found that with the growing number of agents, this average is also growing, which means that the output length is growing faster than linear. This contributes to the fact that the more agents, the more complex the environment, and thus, the longer explanation and more iterations until a correct plan is needed. We also observe that on failing scenarios where the agents are going back and forth, the token length is generally shorter, given that they did not run into the iterative deconflict process, which takes a lot of tokens to solve.

\end{document}